\documentclass{iau} 
\usepackage{graphicx}
\usepackage{subfigure}

\title[Magnetohydrostatic flux tube model in the solar atmosphere] 
{Flux tube model in the solar atmosphere}

\author[S. Sen \& A. Mangalam]   
{S. Sen$^1$
 \and A. Mangalam$^2$}

\affiliation{Indian Institute of Astrophysics\\ Sarjapur Road, Koramangala, Bangalore, 560034, India, \\ email: {\tt $^1$samrat@iiap.res.in, $^2$mangalam@iiap.res.in}} 

\pubyear{2018}
\volume{340}  
\jname{Long-term datasets for the understanding of solar and stellar magnetic cycles}
\editors{A.C. Editor, B.D. Editor \& C.E. Editor, eds.}
\begin{document}

\maketitle

\begin{abstract}
We construct two classes of the magnetohydrostatic equilibria of the  axisymmetric flux tubes with twisted magnetic fields in the stratified solar atmosphere that span from the photosphere to the transition region. We built the models by incorporating specific forms of the gas pressure and poloidal current in the Grad-Shafranov equation.  This model gives both closed and open field structure of the flux tube. The other open field  model we construct is based on the self-similar formulation, where we have incorporated  specific forms of the gas pressure, poloidal current and two different shape functions. We study the homology of the parameter space that is consistent with the solar atmosphere and find that the estimation of the magnetic structure inside the flux tubes is consistent with the observation and simulation results of the magnetic bright points.        
\end{abstract}
\vspace{-0.2 in}
\firstsection 
\section{Introduction}
The study of the magnetic flux tube is one of the important aspects of the solar physics.  \cite[Sen \& Mangalam (2018)]{Sen_Mangalam2018} (SM18 hereafter) have constructed a closed field model of a flux tube with twisted magnetic field in the stratified solar atmosphere, where a specific form of the gas pressure $p$ and poloidal current $I_p$ are assumed to be the second order polynomial of the flux function, $\Psi$.
Incorporating the forms, ${\displaystyle p=\frac{1}{4\pi}\big(\frac{a}{2}\Psi^2+b \Psi \big)+p_{20}\exp(-2kz)}$ and ${\displaystyle I_p^2=\alpha \Psi^2+2\beta \Psi+I_0^2}$, in the magnetohydrostatic force balance equation, the Grad-Shafranov equation (GSE), where, $a$, $\alpha$, $b$, $\beta$, $I_0$, $k$ and $p_{20}$ are the parameters to be determined by appropriate boundary conditions. The solution of the GSE consists of a homogeneous part $\Psi_h$ and a particular part $\Psi_p$, i.e. $\Psi=\Psi_h+\Psi_p$, where $\Psi_h$ is separable with a Coulomb function in the $r$ direction, and a $z$ part, that decreases exponentially with $z$ is obtained by SM18, whereas, $\Psi_p$ is given by \cite[Atanasiu et al. (2004)]{Atanasiu_etal2004}. Considering the presence of the sheet current at the boundary of the flux tube and using the standard boundary conditions: ${\displaystyle B_r(R,z)=0}$, ${\displaystyle p(R,0)=p_0}$ and ${\displaystyle p(R,z_{tr})=p_{tr}}$, where $R$ denotes the radius of the flux tube, $B_0$ is the field strength at the center of the flux tube, $p_0=1.22 \times 10^5$ dyne cm$^{-2}$ and $p_{tr}=0.148$ dyne cm$^{-2}$ are the gas pressures at the photosphere ($z=0$) and the transition region ($z=2$ Mm) respectively [\cite[Avrett \& Loeser (2008)]{Avrett_Loeser2008}], we obtain the magnetic and thermodynamic structures inside the flux tube in terms of the free parameters $(R, B_0)$. The general solution: $\Psi=\Psi_h+\Psi_p$, gives an open field structure, whereas, $\Psi=\Psi_h$, gives a closed field structure, which is discussed in SM18. The other model we built is associated with the self-similar formulation, where, the coordinates $r$ and $z$ are combined together into a  self-similar parameter $\xi$, which leads to the condition, $\Psi(r,z)=\Psi(\xi)$ where, $\xi= \zeta(z) r$. Incorporating a specific form of the gas pressure ${\displaystyle p=p_c \exp(-2kz)+\frac{f}{2} \Psi^2}$, poloidal current $I_p^2=\chi \xi^4 D^2(\xi)$, into the magnetohydrostatic force balance equation and using the boundary conditions mentioned above, we obtain an open field flux tube solution in terms of the homology parameters $\Psi_b,\ B_0,\ B'_{z0},\ p_c$ and $\chi$, where $\Psi_b$ and $B'_{z0}$ denote the boundary flux and vertical field gradient at the photosphere respectively. We use two different profiles for $D$ which are, generalized Gaussian, ${\displaystyle D_G(\xi)=\Psi_b \exp(-\xi^{n_g})}$ and power law function, ${\displaystyle D_P(\xi)=\Psi_b (1+\xi)^{-n_p}}$, and obtain the magnetic and thermodynamic structures inside the flux tube in terms of the homology parameters $\Psi_b,\ B_0,\ B'_{z0},\ p_c$ and $\chi$, which are taken as the consistent values with the observations. 
\begin{figure}
\begin{center}
\subfigure[]{\includegraphics[scale=0.19]{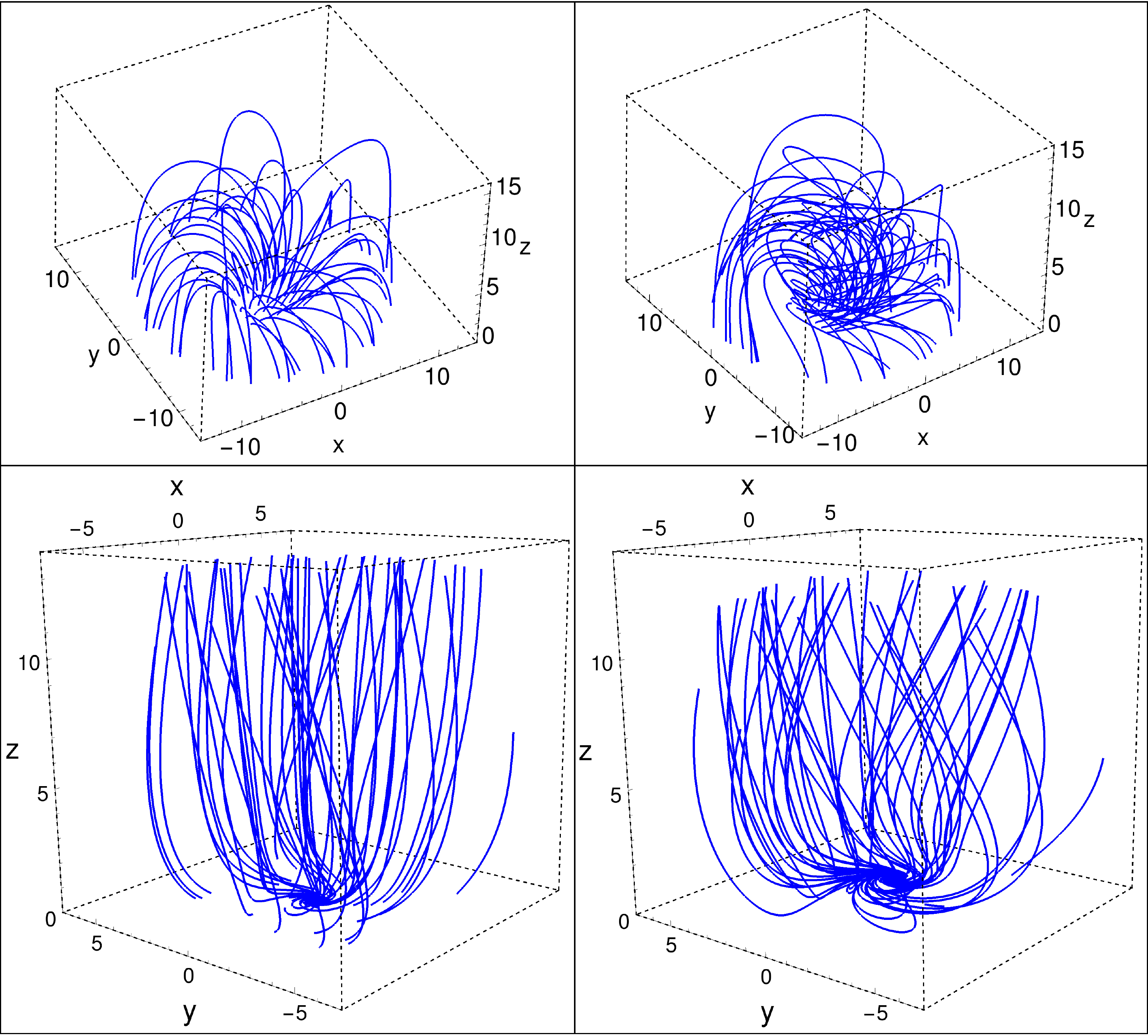}}\hspace{0.15 in}
\subfigure[]{\includegraphics[scale=0.19]{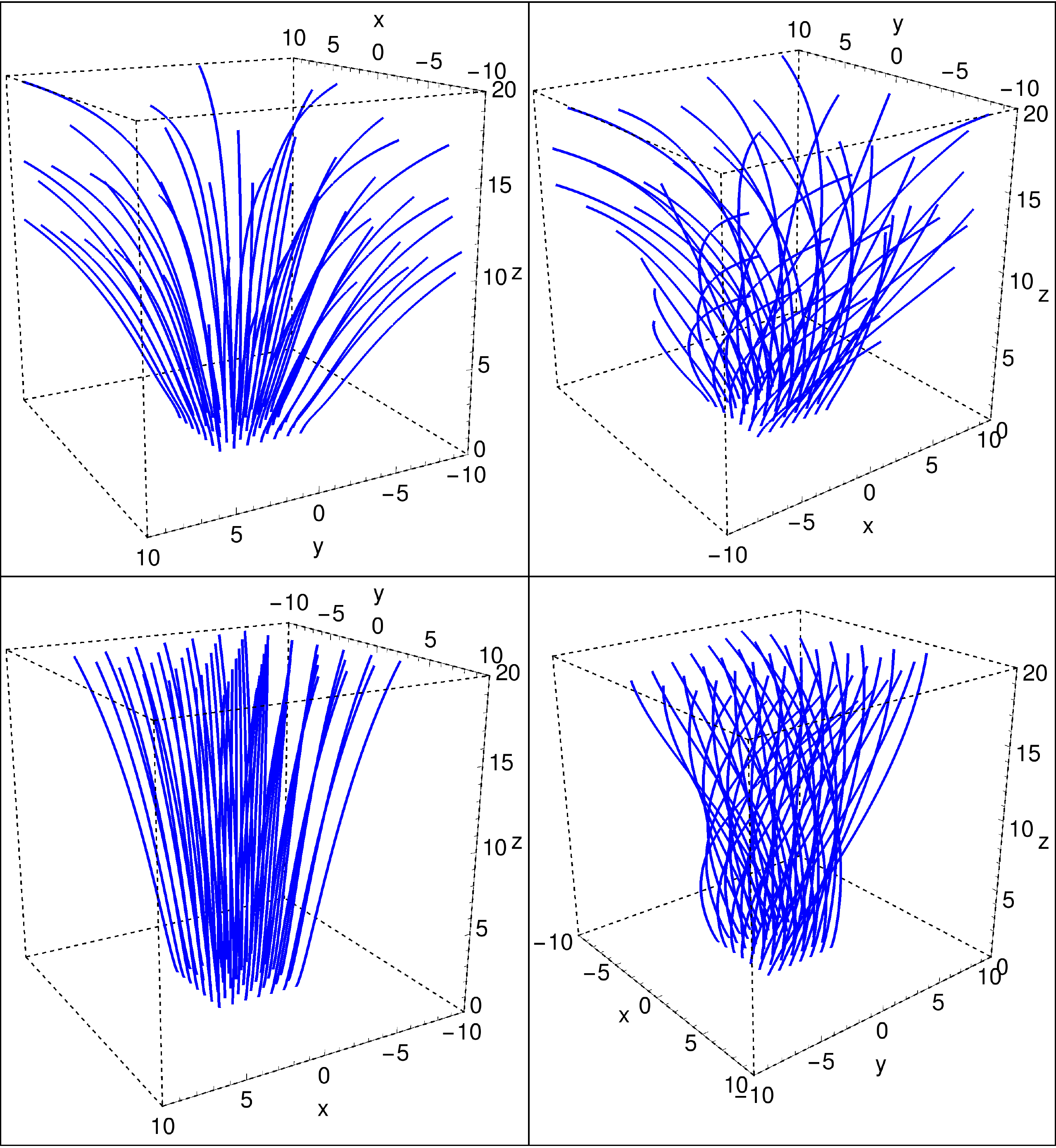}}
\caption{The 3D configuration of the magnetic field lines inside the flux tubes. \textit{Left:} (a) the closed and open field Coulomb function models are shown in the upper and lower panels respectively. \textit{Right:} (b) the self-similar flux tube models are shown in the upper and lower panels for Gaussian ($n_g=2$) and power law (with $n_p=3$) profiles respectively.}
\label{flplots.pdf}
\end{center}
\end{figure}
The 3D field geometry inside the flux tubes are shown in the Figure \ref{flplots.pdf} for both Coulomb function and self-similar models. The more detailed study of the models is discussed in the paper in preparation by Sen \& Mangalam (2018b).
\vspace{-0.24 in}
\section{Conclusions}
We have obtained two classes of the flux tube models with the twisted field which are Coulomb function and self-similar models. The Coulomb function model gives both closed and open field structure, whereas the self-similar model gives an open field structure of the flux tube. The parameters for the Coulomb function model are $[ B_0 =1$ kG$, R=120$ km (left panels of Figure \ref{flplots.pdf}(a)), $R=140$ km (right panels of Figure \ref{flplots.pdf}(a))$]$, and for the self-similar model, $[ \Psi_b=10^{17}$ Mx, $B_0=1$ kG, $B'_{z0}=1$ G km$^{-1}$, $p_c=10^5$ dyne cm$^{-2}$, $\chi=10^{-16}$ cm$^{-2}$ (left panels of Figure \ref{flplots.pdf}(b)), $\chi=10^{-14}$ cm$^{-2}$ (right panels of Figure \ref{flplots.pdf}(b))$]$, which are consistent with the observational values for the magnetic bright points observed in the solar photosphere. The obtained magnetic structure inside the flux tube for both the models are in good agreement with other simulation results e.g. \cite[Gent et al. (2013)]{Gent_etal2013} and \cite[Shelyag et al. (2010)]{Shelyag_etal2010}.

\end{document}